\documentclass[conference]{IEEEtran}
\IEEEoverridecommandlockouts

\ifCLASSINFOpdf
\else
\fi

\usepackage{balance}
\usepackage{setspace, amsmath, amssymb, url, lscape, subfig, algorithmic, multirow, pslatex, listings, verbatim, alltt, amsfonts, wrapfig, boxedminipage, color, cite}
\usepackage[vlined,linesnumbered,ruled,boxed]{algorithm2e}
\usepackage[dvips]{graphicx}
\usepackage{epsfig}

\usepackage{xcolor}
\usepackage{nicematrix}
\usepackage[colorlinks, citecolor=blue]{hyperref}
\usepackage{array}
\usepackage{tikz}

\usepackage{amssymb}
\usepackage{pifont}
\usetikzlibrary{arrows,shapes,positioning}

\usepackage{ragged2e}

\newcommand{\qed}{\nobreak \ifvmode \relax \else
      \ifdim\lastskip<1.5em \hskip-\lastskip
     \hskip1.5em plus0em minus0.5em \fi \nobreak
      \vrule height0.75em width0.5em depth0.25em\fi}

\newcommand{\eg}{{\it e.g., }}
\newcommand{\etal}{{\it et~al., }}
\newcommand{\ie}{{\it i.e., }}
\newcommand{\name}{FastFreeze Daemon\xspace}
\newcommand{\fastmig}{FastMig\xspace}

\newcommand{\comments}[1]{}
\newcommand\hl{\bgroup\markoverwith
  {\textcolor{yellow}{\rule[-.5ex]{2pt}{2.5ex}}}\ULon}

\newcommand{\ff}{FastFreeze\xspace}
\newcommand{\pid}{PID\xspace}
\newcommand{\nslpid}{\texttt{ns\_last\_pid}\xspace}
\newcommand{\nslpidpath}{\texttt{/proc/sys/kernel/ns\_last\_pid}\xspace}

\newcommand{\memhog}{\texttt{memhog}\xspace}
\newcommand{\yolo}{\texttt{YOLOv3-tiny}\xspace}
\newcommand{\revised}[1]{{#1}} 

\lstdefinestyle{mystyle}{
    basicstyle=\ttfamily\footnotesize,
    breakatwhitespace=false,         
    breaklines=true,                 
    captionpos=b,                    
    keepspaces=true,                 
    numbers=left,                    
    numbersep=5pt,            
    xleftmargin=2em,
    showspaces=false,                
    showstringspaces=false,
    showtabs=false,                  
    tabsize=2,
    frame=single,
    framexleftmargin=1.5em,
}

\newcolumntype{C}[1]{>{\centering\arraybackslash}m{#1\textwidth}}

\graphicspath{ {./Figures} }
\def\BibTeX{{\rm B\kern-.05em{\sc i\kern-.025em b}\kern-.08em
    T\kern-.1667em\lower.7ex\hbox{E}\kern-.125emX}}
\begin{document}

\title{\fastmig: Leveraging \ff to Establish Robust Service Liquidity in Cloud 2.0\\

}


\author{
    \IEEEauthorblockN{Sorawit Manatura}
    \IEEEauthorblockA{
        HPCNC Lab, Department of Computer Engineering\\
        Kasetsart University,
        Thailand\\
        sorawit.man@ku.th
    }\\
    \IEEEauthorblockN{Chantana Chantrapornchai*}
    \IEEEauthorblockA{
        HPCNC Lab, Department of Computer Engineering\\
        Kasetsart University,
        Thailand\\
        fengcnc@ku.ac.th
    }
    \and
    \IEEEauthorblockN{Thanawat Chanikaphon}
    \IEEEauthorblockA{
        HPCC Lab, School of Computing and Informatics\\
        University of Louisiana at Lafayette,
        LA, USA\\
    thanawat.chanikaphon1@louisiana.edu
    }\\
    \IEEEauthorblockN{Mohsen Amini Salehi*}
    \IEEEauthorblockA{
        HPCC Lab, Computer Science and Engineering Department\\
        University of North Texas,
        TX, USA\\
        mohsen.aminisalehi@unt.edu
    }
}


\IEEEaftertitletext{\vspace{-2\baselineskip}}



\maketitle

\IEEEpeerreviewmaketitle
\IEEEaftertitletext{\vspace{-10\baselineskip}}
\begin{abstract}
Service liquidity across edge-to-cloud or multi-cloud will serve as the cornerstone of the next generation of cloud computing systems (Cloud 2.0). Provided that cloud-based services are predominantly containerized, an efficient and robust live container migration solution is required to accomplish service liquidity. In a nod to this growing requirement, in this research, we leverage
\emph{\ff}, a popular platform for process checkpoint/restore within a container, and promote it to be a robust solution for end-to-end live migration of containerized services. In particular, we develop a new platform, called \fastmig that proactively controls the checkpoint/restore operations of \ff, thereby, allowing for robust live migration of containerized services via standard HTTP interfaces. The proposed platform introduces post-checkpointing and pre-restoration operations to enhance migration robustness. Notably, the pre-restoration operation includes containerized service startup options, enabling warm restoration and reducing the migration downtime. In addition, we develop a method to make \ff robust against failures that commonly happen during the migration and even during the normal operation of a containerized service. Experimental results under real-world settings show that the migration downtime of a containerized service can be reduced by 30X compared to the situation where the original \ff was deployed for the migration. Moreover, we demonstrate that \fastmig and warm restoration method together can significantly mitigate the container startup overhead. Importantly, these improvements are achieved without any significant performance reduction and only incurs a small resource usage overhead, compared to the bare (\ie non-\ff) containerized services.
\end{abstract}


\section{Introduction}\label{sec:introduction}
\subsection{Motivation}
Cloud 2.0 marks the next generation in cloud computing, integrating cutting-edge technologies like AI, machine learning, and IoT. This phase emphasizes greater efficiency, scalability, and the use of heterogeneous resources, setting the stage for advanced and innovative applications.
Services are envisaged to be truly distributed, either across edge-to-cloud---to support low-latency services \cite{mancuso2023efficiency}---or across multi-cloud---to attain high reliability and cost efficiency \cite{imran2020multi}. The next generation of services in Cloud 2.0 desire \emph{service liquidity} across the resource continuum. That is, the services must be free to execute and relocate seamlessly over distributed and heterogeneous platforms \cite{liqo}. In addition to the futuristic use cases, service liquidity offers several advantages for day-to-day IT operations, such as facilitating maintenance tasks (\eg performing hardware upgrades) and improving resource utilization via dynamically reallocating workloads across hosts without disrupting their operations. Establishing service liquidity entails a seamless live service migration solution from one host to another without degrading the quality of experience.

\begin{figure}[htpb]
    \centerline{\includegraphics[width=0.49\textwidth]{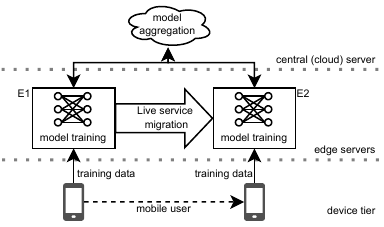}}
    \caption{Service liquidity use case to efficiently achieve federated learning for mobile users. The model training process can seamlessly migrate from a source edge sever (E1) to a destination one (E2). }
    \label{fig:use_case}
\end{figure}

As an exemplar use case, consider the federated learning scenario, shown in Figure~\ref{fig:use_case}, where the model training takes place only on the on-premises edge servers, to preserve the user's data privacy. Then, the trained model is transferred to the central (cloud) server to perform model aggregation \cite{ullah2022fedfly}. In this scenario, the data collection is often performed on the device (\eg smartphone) of a mobile user who tends to randomly disconnect from one edge server and connect to another. Such mobility can slow down the training if the model has to be retrained. The ability to live migrate the training service across edge servers, however, enables seamless resumption of the training process at the new edge server, thereby, saving the training time \cite{ullah2022fedfly}.

In modern cloud environments, where services are increasingly containerized for the merit of portability and isolation, various migration approaches are studied, including the native approach that requires the container runtime checkpoint/restore feature \cite{stoyanov2018efficient} and the container nesting approach \cite{chanikaphon2023ums} that nests the containerized service within another container providing the migration ability. A distinct approach that aims not to migrate an entire container but to migrate only the ``embedded processes of the service'' is desired to minimize the overhead, maximize portability, and cater to multi-cloud benefits. \ff\cite{FastFreeze} is a turn-key solution that exhibits these features; however, it is purposed for containerized processes checkpointing and restoration.
\ff packages up the necessary dependencies into a single library for building a container image, making it easy-to-use through a simple command-line interface and unprivileged to be able to execute inside a container securely. In this work, our hypothesis is that \ff can be an ideal candidate to implement live migration of containerized services and realize the idea of liquid computing.

\subsection{Problem Statement}\label{subsec:problem}
\begin{figure}[htpb]
  \centering
    \subfloat[Traditional container migration]{\includegraphics[width=0.49\textwidth]{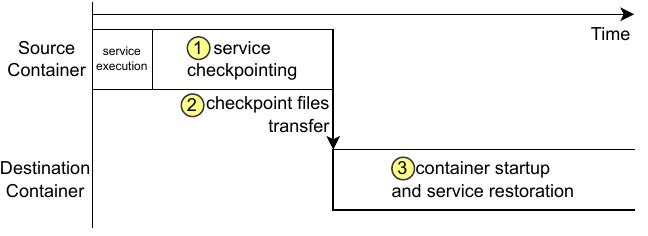}\label{fig:post_checkpoint_pre_restore_traditional}}
  \hfill
    \subfloat[Proposed container migration with post-checkpointing and pre-restoration steps]{\includegraphics[width=0.49\textwidth]{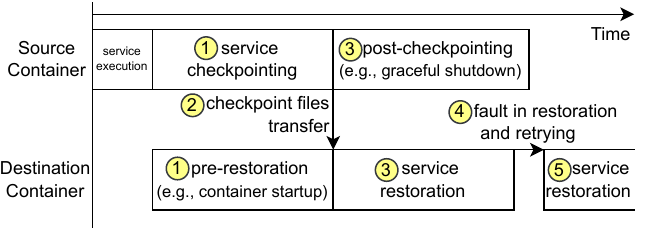}\label{fig:post_checkpoint_pre_restore_proposed}}
    \caption{Positioning of the post-checkpointing and pre-restoration steps in the live container migration process. The traditional process includes three main steps: \textcircled{\raisebox{-.8pt} {1}} service checkpointing, \textcircled{\raisebox{-.8pt} {2}} checkpoint files transfer, and \textcircled{\raisebox{-.8pt} {3}} new container startup and service restoration. The proposed solution incorporates the pre-restoration operations that execute simultaneously in step \textcircled{\raisebox{-.8pt} {1}} and, similarly, post-checkpointing operations in step \textcircled{\raisebox{-.8pt} {3}}. Decoupling the container startup from the service startup is the key to achieving fault tolerance in steps \textcircled{\raisebox{-.8pt} {4}} and \textcircled{\raisebox{-.8pt} {5}}.}
    \label{fig:post_checkpoint_pre_restore}
\end{figure}

To realize the live containerized service migration, developers may choose to simply import an existing checkpoint/restore solution, like \ff, into the container image. Nevertheless, our preliminary analysis shows that this approach is failure-prone due to the design of a modern container and the uncertainties of the migration process across systems.
Figure~\ref{fig:post_checkpoint_pre_restore_traditional} depicts the necessary steps for live migration: service checkpointing, checkpoint files transfer, and restoration. \revised{In the traditional container, the service processes exit after the checkpointing step is completed. Then, the entire container is considered down by the runtime and loses a chance to execute graceful shutdown instructions, which can adversely affect the source system. For instance, the exit behavior of the services may differ from what is expected, potentially disrupting subsequent operations at the source system.
Additionally, binding the containerized service management and its execution---or, in other words, binding container startup with the containerized service startup---during the migration limits the containerized service from being \emph{warm}-restored. That is, the container startup instructions must wait unnecessarily for the checkpoint files transfer step to complete.} As a result, we realized that the lack of two additional steps---post-checkpointing and pre-restoration, shown in Figure~\ref{fig:post_checkpoint_pre_restore_proposed}---hinders the establishment of a robust container migration service.

In this paper, we adopt \ff to be the underlying vehicle for a sustainable containerized service migration. We propose the \fastmig container that is equipped with the service management layer, as shown in Figure~\ref{fig:overview}, that enables a fast and robust service migration across computing systems. Decoupling the service management layer from the service execution layer enables the pre-restoration operations, that reduce the service migration time, and the post-checkpointing operations that pave the way to execute graceful shutdown instructions, thus enhancing the service robustness.

\revised{
However, it does not fully establish robust service liquidity as \fastmig is still prone to faults that occur randomly---both in normal operations and in the container migration process---and cease the running service processes. Traditional fault tolerance mechanisms that offer container recreation are unsuitable for \fastmig as the container is reusable for multiple startups/restorations of the same service; hence, they fail to maximize restart efficiency---by performing unnecessary container startup \cite{denninnart2024smse}---and traceability---by segregating service execution logs to multiple containers.
The proposed \fastmig service management layer includes a fault tolerance mechanism that allows restarting the service without recreating the container. As a result, it is capable of handling faults in an efficient manner and satisfies service liquidity robustness.
}

Lastly, \fastmig promotes \ff to be migration-friendly. The problem is that \ff imposes an unusual latency on the migration of a containerized service that forks multiple child processes, particularly, at the restoration time. \fastmig, however, improves the performance of the restore operation for multi-process services and reduces the service migration time significantly. 
\fastmig also offers a standard interface to securely serve external migration requests.

\subsection{Contributions}
In sum, this paper makes the following contributions:
\begin{itemize}
\item Developing \fastmig, a live containerized service migration solution that incorporates \ff with the service management layer to realize service liquidity
\footnote{\fastmig and all the experimental data are publicly available for reproducibility purposes in the following addresses: \url{https://github.com/hpcclab/fastfreeze4service_migration}}
\item Developing \name that decouples the containerized service management from its execution. It enables post-checkpointing and pre-restoration operations and incorporates the warm restoration technique that reduces the migration time.
\item Developing a configurable fault tolerance mechanism that enhances service liquidity robustness via allowing restarting the service without recreating the container.
\item Extending \ff to improve the restoration time of the multi-process services upon migration and develop standard APIs to make \fastmig pluggable to other solutions.
\item Evaluating and analyzing the impact of \fastmig under real-world scenarios and settings.
\end{itemize}

The rest of this paper is organized as follows: Section \ref{sec:background} provides background on the live containerized service migration and \ff.
Section \ref{sec:robust} presents the design of the robust containerized service using \ff.
Section \ref{sec:evaluation} describes the evaluation and the results. Section \ref{sec:related-works} discusses related studies on service liquidity and live containerized service migration.
Finally, Section \ref{sec:conclusion} concludes our work.
\section{Background}
\label{sec:background}
\subsection{Live Containerized Services Migration}
Container technology is a virtualization technology that bundles the application and all its dependencies, providing application isolation and making the application migratable while retaining complete functionality \cite{docker-container,redhat-containers}. The ability to migrate containerized services between systems is beneficial in service provisioning through load balancing or fault tolerance. On a high level, for live migration, the containerized service migration can be done by checkpointing the processes inside the container, moving the checkpoint files to the destination, and restoring the container to its original state \cite{stoyanov2018efficient}. There are techniques to help reduce the downtime that comes from the transfer process, such as pre/post-copy \cite{singh2021taxonomy}  or page-server \cite{stoyanov2018efficient}, but the overall operation remains the same. A main tool that has been used to do the container checkpoint/restore operations is Checkpoint/Restore In Userspace (CRIU) \cite{CRIU}. CRIU achieves this by using \texttt{ptrace}, a kernel interface to inspect the current process execution and memories used. CRIU also works with multi-process applications as it can checkpoint and restore an entire process tree. 

When using CRIU to restore an application, the \pid{}s of the application processes have to be the same as before it was checkpointed. With enough privileges or permission (\eg \texttt{root} or \texttt{CAP\_SYS\_ADMIN} capability\footnote{\revised{Capability is a privilege for specific functionality in Linux, \eg the ability to change file owner. In analogy, it is a subset of root privileges that can be granted to unprivileged processes. To minimize the attack surface, it is a practice to grant only the capabilities the process requires to function.}}), CRIU can achieve the desired \pid at restoration time. Another solution to control the processes \pid is to modify the file \nslpidpath, which indicates the last \pid allocated and determines the next fork \pid in the current PID namespace\footnote{\revised{PID namespace is a Linux kernel feature that isolates process ID number space. A process in an isolated namespace can have an arbitrary PID independent from existing ones in other namespaces. Processes are not visible across namespaces except in a number of cases. PID namespace is one of the important foundations of container technology.}}, to match its need (\eg desired \pid-1).

\subsection{\ff}
\label{subsec:bgFF}
FastFreeze is a checkpoint/restore utility built specially for containerized services. It is a wrapper around CRIU; hence, \ff delegates the main processes checkpoint/restore to CRIU. \ff provides management facilities for checkpoint/restore, providing a customized \emph{init} process\footnote{\revised{The container \emph{init} process is the root of the container process tree. Its status represents the container status; for example, the container is considered down if its \emph{init} process exits. The \emph{init} process is also responsible for other duties, such as forwarding signals to child processes and reaping zombie processes.}} that makes the containerized service processes its children. 

In addition, \ff is designed to be usable within unprivileged containers. It uses a technique called \textit{fork bomb}, rapidly forking and killing child processes until it gets the desired \pid to handle the CRIU \pid requirement upon restoration.
For single-process application restoration, the delay is not significant; however, for multi-process applications \cite{ghatrehsamani2020art}, the delay from the \textit{fork bomb} procedure can be significant. Chanikaphon \etal \cite{chanikaphon2023ums} show that the migration approach using \ff has increased the migration overhead by $\approx$40 seconds per 1 additional child process. 

Lastly, \ff provided only a command-line interface to activate all its operations; this leads to obstacles in the migration requirements in the aspects of container access permission, synchronization, and monitoring. Moreover, the lack of easy-to-use/standard interfaces obstructs the integration with web services and leads to incompatibility with service-oriented architecture (SOA). 

\section{\ff for Robust Service Liquidity across Computing Systems}
\label{sec:robust}
\begin{figure}[htpb]
    \centerline{\includegraphics[width=0.49\textwidth]{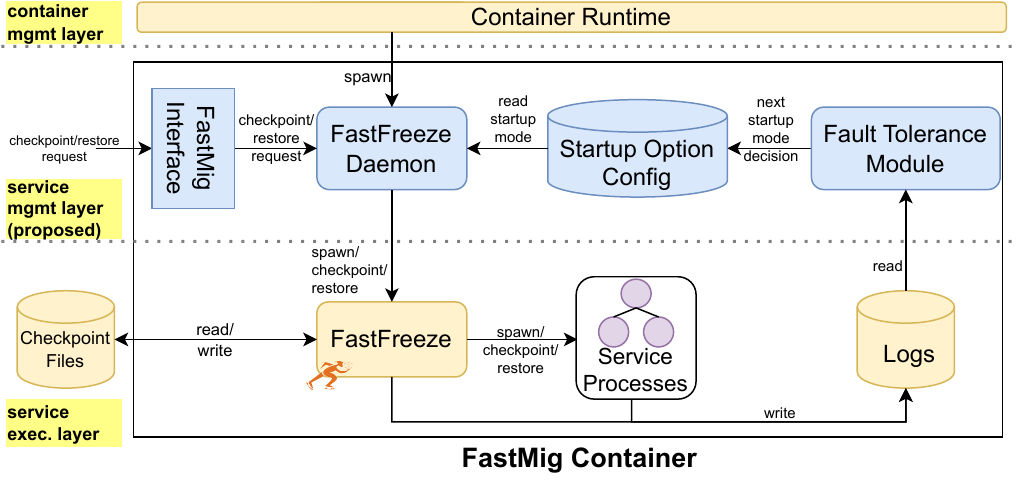}}
    \caption{Overview of the \fastmig within a container. We propose adding the ``service management layer'' (components with the blue color) to enable fast and robust live migration of containerized services.}
    \label{fig:overview}
\end{figure}
\subsection{Overview}
Figure~\ref{fig:overview} illustrates the overview of \fastmig, the system that utilized the adapted \ff for robust service liquidity across computing systems. The system consists of 5 main modules: (A) \emph{the \fastmig Interface} allowing requests from outside, (B) \emph{\name}, providing containerized service management, (C) \emph{\ff}, utilizing CRIU for checkpoint/restore operations and acting as parent process of the containerized service, (D) \emph{the containerized service} and, (E) \emph{the Fault Handling Module}, which is the configurable logic used to handle the fault from logs. Figure \ref{fig:overview} also shows another component called \emph{the Start Option Config}, which is the file that indicates how the service will be started.

\subsection{Adapting \ff to Establish Robustness}

To enhance robustness, we address three key aspects. Firstly, we separate containerized service management and its execution and examine startup options, enabling finer service control during migration. Secondly, we establish a fault tolerance mechanism to bolster the resilience of containerized services during both regular operation and migration. Lastly, we offer the \fastmig Interface to facilitate seamless integration with migration systems, enhancing system security.

\subsubsection{Service Management Decoupling}

For modern containerized services, the container and the containerized service execution are bound together. For instance, when the service exits, the container also exits, and the service is started upon the container start. This factor limits the ability to manage the containerized service lifetime during the migration, as discussed in Section~\ref{subsec:problem}.
To establish the robustness of service liquidity, we build a solution that allows us to separate the containerized service management (\eg states) and its execution. \ff acts as a parent process of the containerized service, covering the service execution layer; therefore, we developed a component for the service management layer called \name.

\begin{figure}[htpb]
  \centering
    \subfloat[Traditional container]{\includegraphics[width=0.49\textwidth]{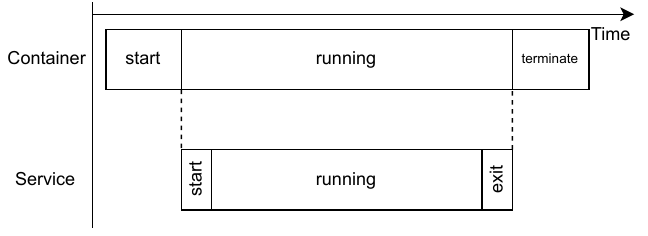}}
  \hfill
    \subfloat[\fastmig container]{\includegraphics[width=0.49\textwidth]{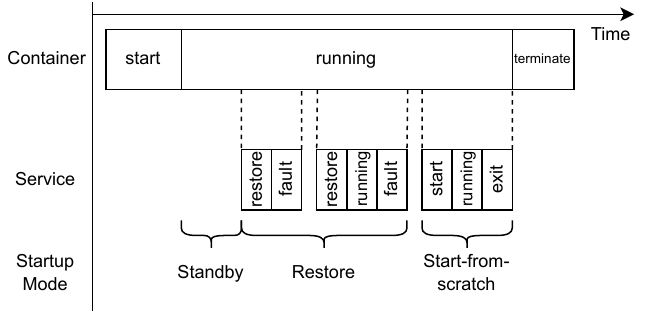}}
    \caption{\revised{Container and containerized service lifespan in traditional and \fastmig containers. \fastmig container enhances robustness by allowing the service to restart without recreating a new container. For each restart, the fault tolerance mechanism determines how the containerized service starts: restore from checkpoint files, start from scratch, or standby (not starting).}}
    \label{fig:startup_modes}
\end{figure}

\name provides a standby behavior for the containerized services. The container can be in a running state while the internal service is not running. As \name is the first and always running process in the container, in standby mode, it listens for the incoming request to run the service up, either restore or start it from scratch.
The containerized services equipped with \name can start in three different ways, depicted in Figure~\ref{fig:startup_modes}: (i) start the service from scratch disregarding the checkpoint; (ii) restore the service from the given checkpoint; or (iii) remain on standby---not running the service. The startup behavior for each circumstance is configurable by providing \ff with the \emph{Startup Option Config} indicating the startup mode and relevant information. If no configuration is provided, \name remains in the standby mode by default.
A notable use case for the standby mode is that in the migration process, one can start the container of a service at the destination using the standby mode. When the checkpoint files are transferred, the service can then restore with the container already warm-up. We named this technique \emph{warm restoration}, inspired by a serverless function warm start \cite{denninnart2023efficiency,warmstart}.

\revised{The separation of the service management layer and the service execution layer, together with startup modes, enhances the service migration robustness by allowing post-checkpointing/pre-restoration operations such as graceful shutdown and warm restoration. Moreover, it enables restarting the service when faults occur without creating a new container via the fault tolerance mechanism. Not only does it eliminate unnecessary startup overhead for a new container, but it also enhances service traceability by eliminating the need to aggregate logs across old and new containers.}

\subsubsection{Fault Tolerance Mechanism}
\label{subsub:fault}
Traditional \ff already implemented a feature for users (\ie service developers) to plug in the metric recorder program, which consumes \ff JSON-structured logs as its argument. We developed the \emph{Fault Handling Module} that analyzes logs of how the service exits and decides how it should restart. After that, it outputs its decision into the \emph{Startup Option Config}, and the \name then reads it in the next service startup. In summary, we use an analogy to a self-feedback method.

We defined a default logic that can handle simple faults based on the service exit code as a demonstration. The pseudo-code of the mentioned logic is shown in Listing \ref{lst:default_fault}. The logic indicates that when the service exits as expected with code 0, \name will try to restore the service from the checkpoint files in the next startup, and if the service exits by receiving signals, including the signal from \ff checkpoint operation, the startup option will be standby; otherwise, \name will force the service to start from scratch.

\begin{lstlisting}[language=Python,caption=Default Fault Handling Module logic written in Python language, label={lst:default_fault}]
if exit_code == 0:
    start_option = restore
elif exit_code >= 128 and exit_code <= 159:
    start_option = standby
else:
    start_option = from_scratch
\end{lstlisting}

In other use cases, with different requirements, users can specify their logic for how their containerized services should restart after specific faults and situations happen. 

\subsubsection{The \fastmig Interface}
The \fastmig Interface exposes the \ff operation to be usable from outside of the container. In detail, the HTTP interface is implemented within the \name.
With provided HTTP APIs, external entities can call the migration commands by sending requests, and as distinct from command-line execution, the HTTP request does not require container access permission. As a result, this enhances \ff integration with other systems and enhances the migration operation security.
We achieved this by modifying the \ff to communicate through the Unix domain socket after each operation. \fastmig keeps listening to the socket, waits for the operation finishing sign, and then responds to the client. 
The \fastmig Interface mimics \ff commands for its APIs, so it has two main APIs, namely \texttt{run} API and \texttt{checkpoint} API. Both APIs accept a JSON request body that aligns with each \ff command. After \fastmig Interface receives a request, it processes the body and the Startup Option Config; then it spawns \ff with extracted arguments.

Another significant behavior difference between the HTTP Interface and \ff command-line interface is that the APIs are synchronous, while the prior command-line is not. The \fastmig Interface responds to the client only after the called operation is ended, either finished or failed (\eg the application started successfully). This characteristic makes the checkpoint/restore and migration operations more traceable and eases the evaluation of measuring their duration.

\subsection{Adapting \ff for multi-process Services}
\label{sec:multiprocess}

As mentioned in \ref{subsec:bgFF}, there can be a significant delay when using \ff to restore multi-process services in unprivileged scenarios. This is usually normal and practical cases arising from \ff using the \emph{fork bomb} workaround. 
Our examination reveals that the current \ff solution tries to edit the \nslpidpath first to set the desired process ID, and if it cannot, it will use \emph{fork bomb} until it gets the desired \nslpid, as shown in Listing~\ref{lst:fork_bomb}.

\noindent
\begin{minipage}{\linewidth}
\begin{lstlisting}[caption=The pseudo-code of  Fastfreeze solution to obtain the desired PID during the service restoration, label={lst:fork_bomb}]
set_next_pid(pid):
    if /proc/sys/kernel/ns_last_pid is writable:
        write the desired last_pid(pid-1)
    else:
        use fork_bomb(pid-1)

fork_bomb(pid):
    While ns_last_pid != pid:
        fork_process
        kill_process
    
\end{lstlisting}
\end{minipage}

To avoid the delay, it must prevent the \emph{fork bomb} by allowing \ff to write to \nslpidpath while still not over-giving the privilege to the container and \ff itself. To do that, two limitations prevent \ff from editing  \nslpidpath: (i) without root privilege, \ff has no permission to edit  \nslpidpath, which is a system file, (ii) in the typical container environment(\eg Kubernetes \cite{k8s}, Docker \cite{Docker}), the container mount \texttt{/proc} directory as a read-only directory.

Since Linux kernel version 5.9, it introduced a new capability named \texttt{CAP\_CHECKPOINT\_RESTORE}\cite{CAPCHECKREST}, which provides the ability to control \pid for corresponding PID namespace via editing \nslpid and \texttt{clone3} system call. With this capability, the first limitation can be overcome.
To tackle the second limitation, we study the case of running \ff in a Docker container and adapt the Docker security options \verb|systempath=unconfined| and \verb|apparmor=unconfined|  to allow access to system directories. However, disabling AppArmor allows writing access to all system files and may lead to security issues.\footnote{\revised{AppArmor is a mandatory access control (MAC) system for Linux. It interfaces with the Docker container as another layer of security. The Docker container applies a default AppArmor profile, restricting actions and accesses, \eg avoid editing \texttt{/proc} file system. Instead of just disabling AppArmor, users can provide a secure custom profile for hardening purposes.
}}
Thus, it must be set together with a custom AppArmor profile only to permit modifying \nslpid but not for others in \texttt{/proc} file system.

An example of a Docker configuration avoiding the \emph{fork bomb} delay is shown in Listing \ref{lst:forkbomb} together with an AppArmor profile entry in Listing \ref{lst:apparmor}. The performance improvement is reported in Section~\ref{subsec:fork_bomb_improvement}.

\noindent
\begin{minipage}{\linewidth}
\begin{lstlisting}[caption=Example of Docker run parameters avoiding \ff fork bomb,label={lst:forkbomb}]
Docker run --cap-add=cap_sys_ptrace 
           --cap-add=cap_checkpoint_restore 
           --security-opt systempaths=unconfined 
           --security-opt apparmor=apparmor_config 
           ff_container  
           fastfreeze run app
\end{lstlisting}
\end{minipage}

\noindent
\begin{minipage}{\linewidth}
\begin{lstlisting}[caption=Example of secured AppArmor profile entry,label={lst:apparmor}]
deny @{PROC}/sys/kernel/{?,??,[^s][^h][^m]**}-@{PROC}/sys/kernel/ns_last_pid  w,  
\end{lstlisting}
\end{minipage}
\section{Evaluation}
\label{sec:evaluation}

To evaluate that \fastmig is feasible for live container migration with minimal overheads and downtime while still being robust with the fault tolerance mechanism, we summarized evaluation in a number of aspects in the following metrics:
\begin{enumerate}
    \item The increased overhead on service performance caused by incorporating the \ff and \fastmig into the container
    \item The improvement of the service restoration time when granted appropriate privileges
    \item The improvement of migration time when using \fastmig and warm restoration
    \item The impact of the fault tolerance mechanism on migration performance and its coverage for various types of faults
\end{enumerate}

\subsection{Experimental Setup}
We created Ubuntu 22.04 LTS VMs to represent each as a physical computing node with 4 vCPU, 16 GiB memory, and 100 GiB storage. All VMs are connected with 1 GiB Ethernet. We used Docker version 25.0.1 as the container engine and \ff version 1.3.0.

The live migration performed in the experiments applied the stop-and-copy method by checkpointing the container, dumping checkpoint files to the shared filesystem (NFS), and restoring the container at the destination using the shared checkpoint files.

In most experiments, we deployed 2 distinct applications with different memory footprint behaviors, as it is a critical factor in the migration performance. First, we deployed a popular benchmarking application called \memhog\cite{stoyanov2018efficient}. We configured \memhog to write random data to the
allocated memory and print a counter number every second, representing a static memory footprint application. Second, we configured YOLOv3-tiny\cite{redmon2018yolov3}, a popular object detection application, feeding it with an input image (160KB) from their repository. As opposed to \memhog, \yolo has a dynamic memory footprint.

\begin{table}[htbp]
    \centering
    \begin{tabular}{|c || c | c |} 
 \hline
 Container Types & \%CPU & Mem(MiB)  \\ [0.5ex] 
 \hline\hline
 Bare (non-\ff) & 0.0183	& 128.7 \\ 
 \hline
 \ff & 0.0193 &	129.4  \\
 \hline
 \fastmig & 0.0193 & 130.4  \\
 \hline
 
\end{tabular}
    \caption{Resource Usage}
    \label{tab:1}
\end{table}

\begin{figure}[htpb]
  \centering
    \subfloat[CPU]{\includegraphics[width=0.24\textwidth]{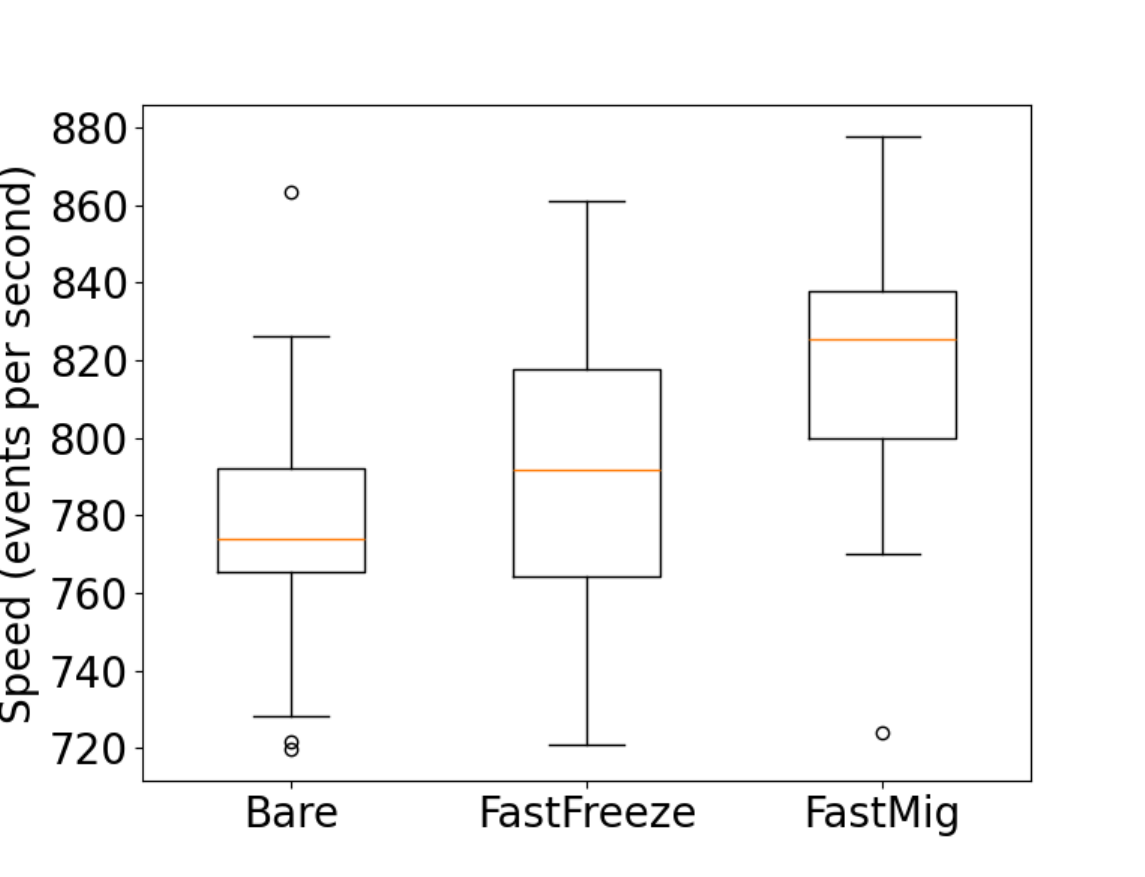}}
  \hfill
    \subfloat[Memory]{\includegraphics[width=0.24\textwidth]{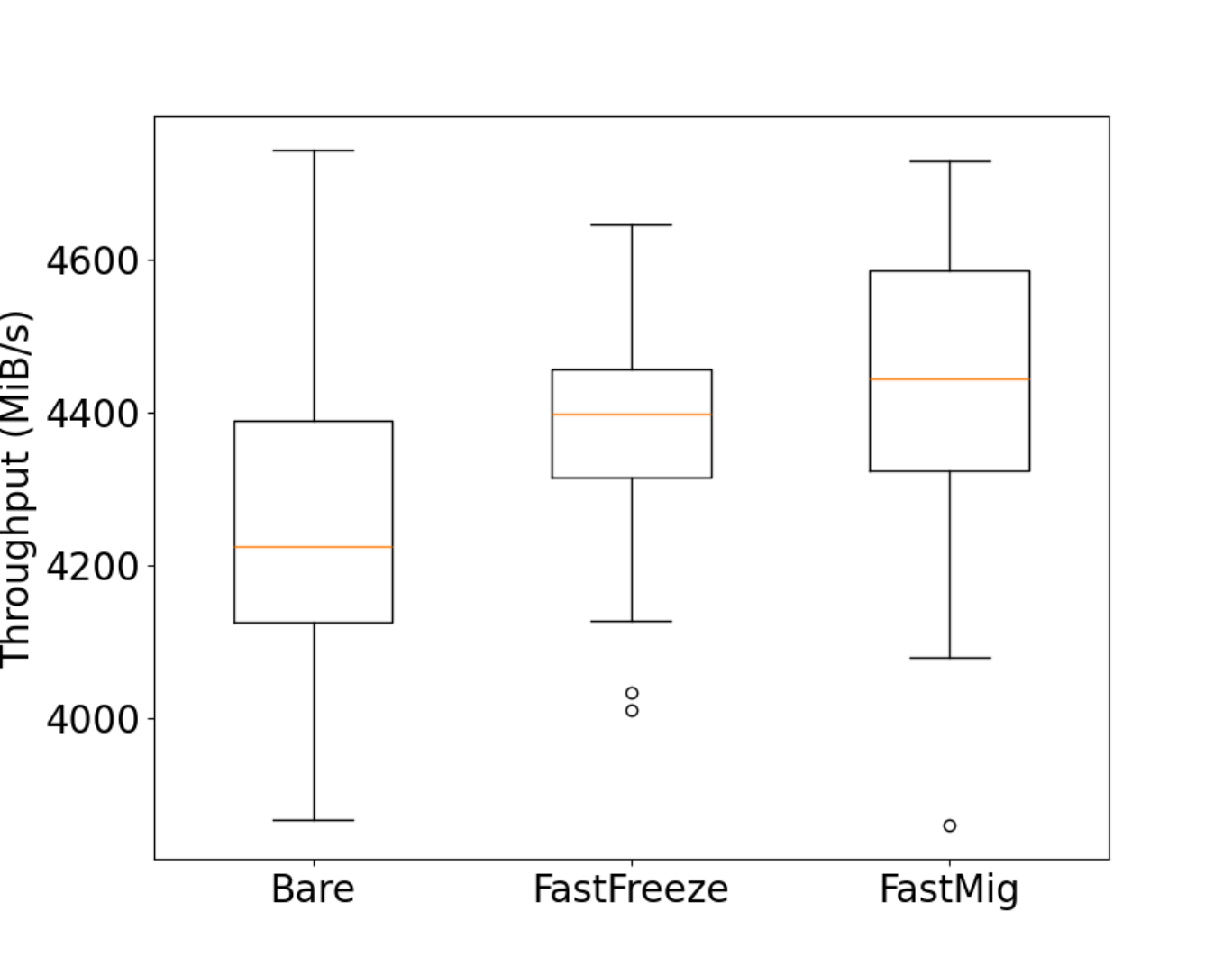}}
  \hfill
    \subfloat[Network]{\includegraphics[width=0.24\textwidth]{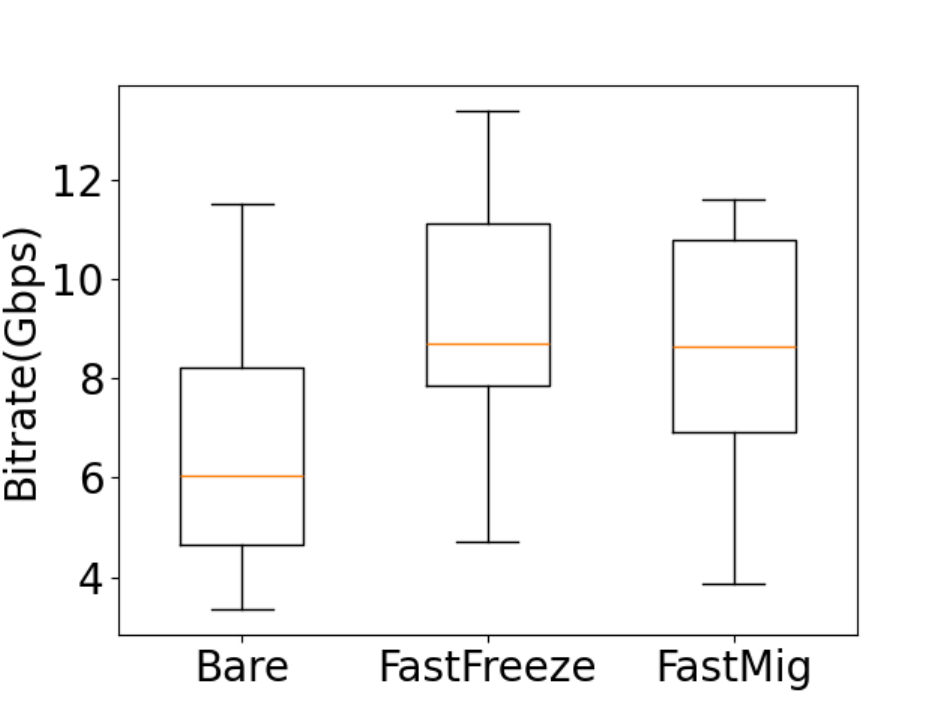}}
  \hfill
    \subfloat[I/O read]{\includegraphics[width=0.24\textwidth]{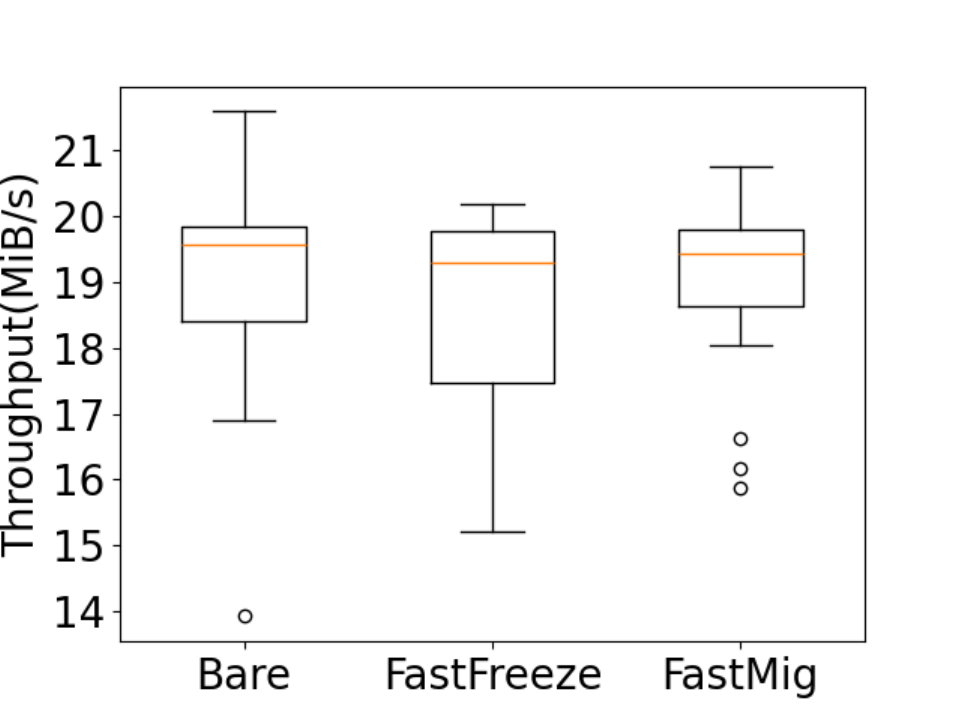}}
  \hfill
    \subfloat[I/O write]{\includegraphics[width=0.24\textwidth]{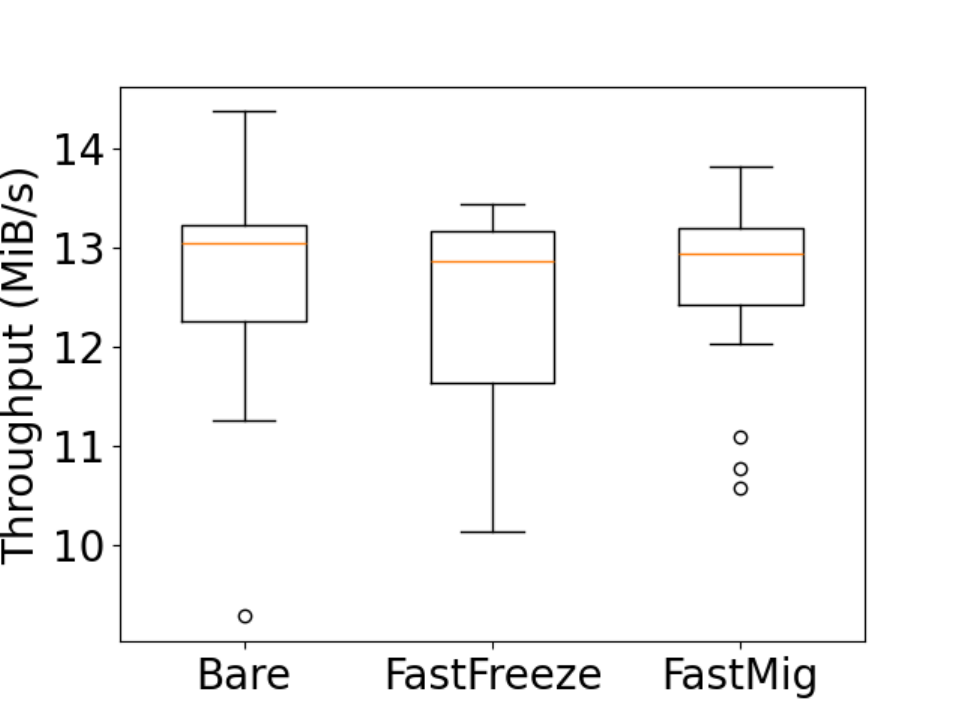}}
    \caption{Performance metrics of regular, \ff-enabled, and \fastmig-enabled service during the normal operation (no migration).
    }
    \label{fig:perf}
\end{figure}

\subsection{Resource Usage and Performance Overhead (No Migration)}
\label{subsec:resource_usage_overhead}
This experiment aims to evaluate the impact of incorporating \ff and \fastmig into the container by measuring service performance during normal operation (\ie, no migration). For that, we configured three kinds of containers, including (A) regular Ubuntu container, (B) Ubuntu container with \ff, and (C) Ubuntu container with \fastmig running, comparing with each other on the container resource usage and performance. 

First, we used the \memhog benchmark, which has a static memory footprint of 128 MiB on each container, to inspect the CPU usage percentage and memory usage with \verb|docker stats| 30 times and calculate the average.
As seen in Table \ref{tab:1}, CPU usage had no significant differences. For memory usage, the container with \ff used $<$1 MiB more than the regular container since \ff will spawn a process (mentioned as customized \emph{init} process in \ref{subsec:bgFF}) and act as the service parent. The memory usage was increased by 1 MiB for the container with \fastmig. Since the average modern container memory footprint is between 50 MB to 300 MB per container \cite{ghatrehsamani2020art}, we conclude that the resource usage overhead is very negligible.

Secondly, we utilized the iperf3 \cite{ermolenko2021internet} benchmark to measure network bitrates and Sysbench \cite{kim2021resource} to measure the performance metrics, including CPU speed (in the number of events processed per second), memory access throughput, and I/O read/write throughput. We ran each benchmark on each container for 30 times. 
The results in Figure~\ref{fig:perf} show that there is no significant performance degradation for all test cases.

\vspace{3mm}
\noindent
\colorbox{blue!10}{
\parbox{0.47\textwidth}{
\textbf{\underline{Takeaway}:} \emph{Enhancing service liquidity by incorporating \ff or \fastmig into the container incurs little to no additional resource and performance overheads.
}
}
}

\begin{figure}[htbp]
    \centerline{\includegraphics[width=0.49\textwidth]{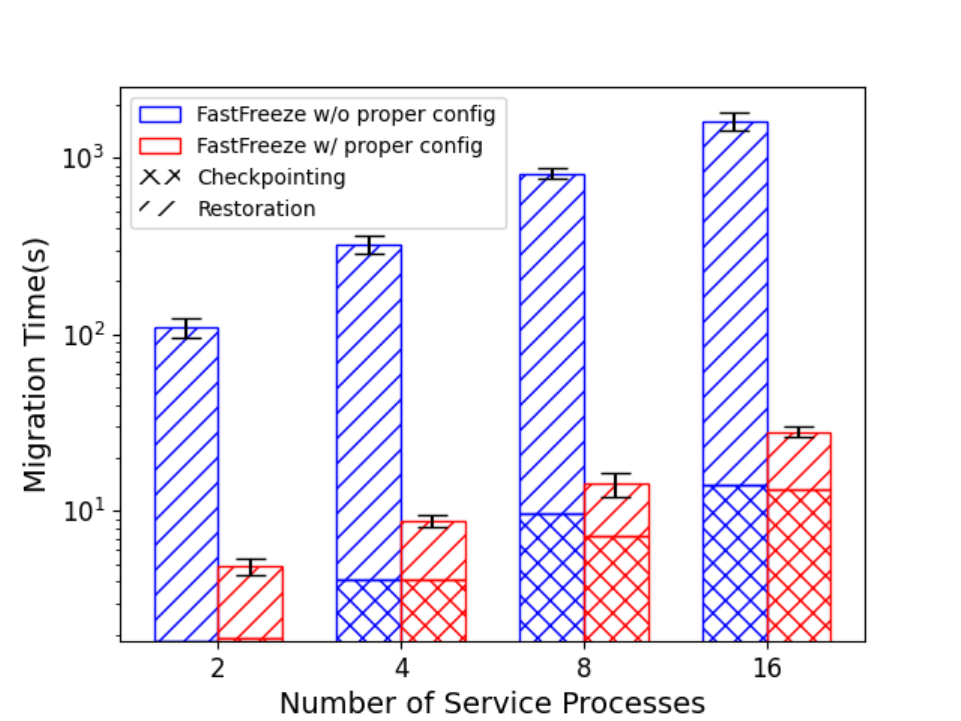}}
    \caption{Comparison of migration time before and after \ff configuration fix. Error bars show 95\% confidence intervals.
    }
    \label{fig:multiprocesses}
\end{figure}

\subsection{Multi-process Application Restoration Time}
\label{subsec:fork_bomb_improvement}
As mentioned in Section~\ref{subsec:bgFF}, \ff restoration time is irregularly high when used with multi-process applications and affects service downtime during migration. The cause examination and fixing were explained in Section~\ref{sec:multiprocess}. We conducted experiments that compared the migration time between using \ff without additional privileges and after allowing it to modify the \nslpidpath. 
We deployed \memhog in a similar manner to the previous experiments in Section~\ref{subsec:resource_usage_overhead}. The number of \memhog processes run in a single container was 2-16, with a scaling factor of 2. Thirty rounds were performed for each case with a different number of processes and reported in Figure~\ref{fig:multiprocesses}.

When allowing \ff to modify the \nslpid file, the migration time was dramatically reduced compared to using unprivileged \ff with the same number of processes. For two processes, the migration time was reduced by $\approx$30 seconds, and for 16 processes, the reduction was greater, \ie $\approx$450 seconds. The differences are mainly influenced by the restoration time differences, as the checkpointing time is mostly identical between cases with the same number of service processes. 

\vspace{3mm}
\noindent
\colorbox{blue!10}{
\parbox{0.47\textwidth}{
\textbf{\underline{Takeaway}:} \emph{To have \ff-based live migration solution operate efficiently with multi-process services, certain privileges must be granted to the container.
}
}
}


\begin{figure}[htpb]
  \centering
    \subfloat[\memhog]{\includegraphics[width=0.49\textwidth]{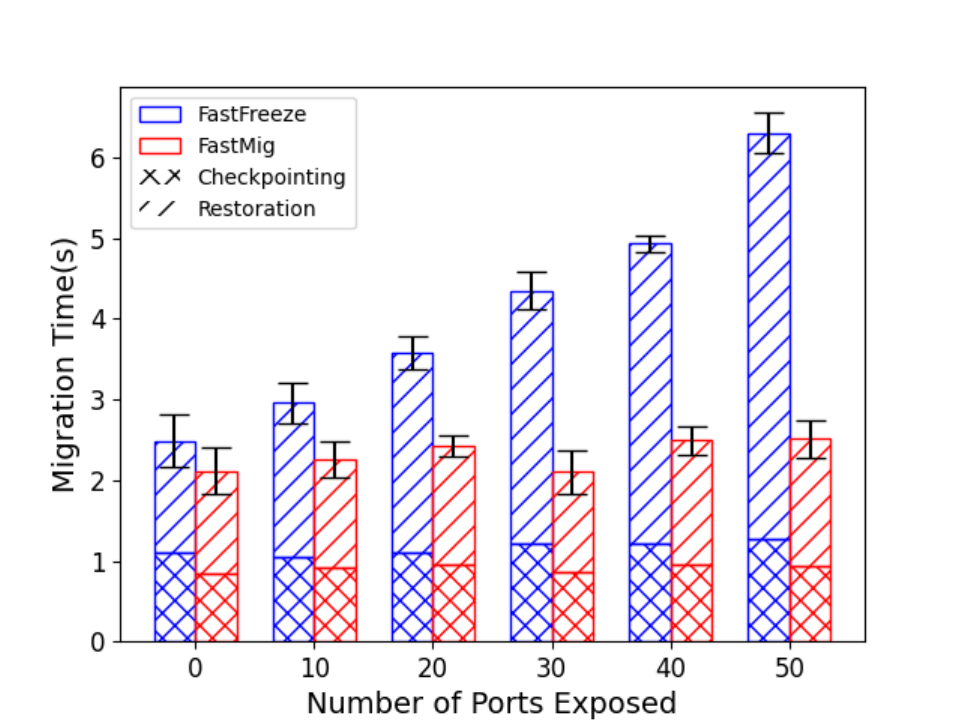}}
    \label{fig:warm_memhog}
  \hfill
    \subfloat[\yolo]{\includegraphics[width=0.49\textwidth]{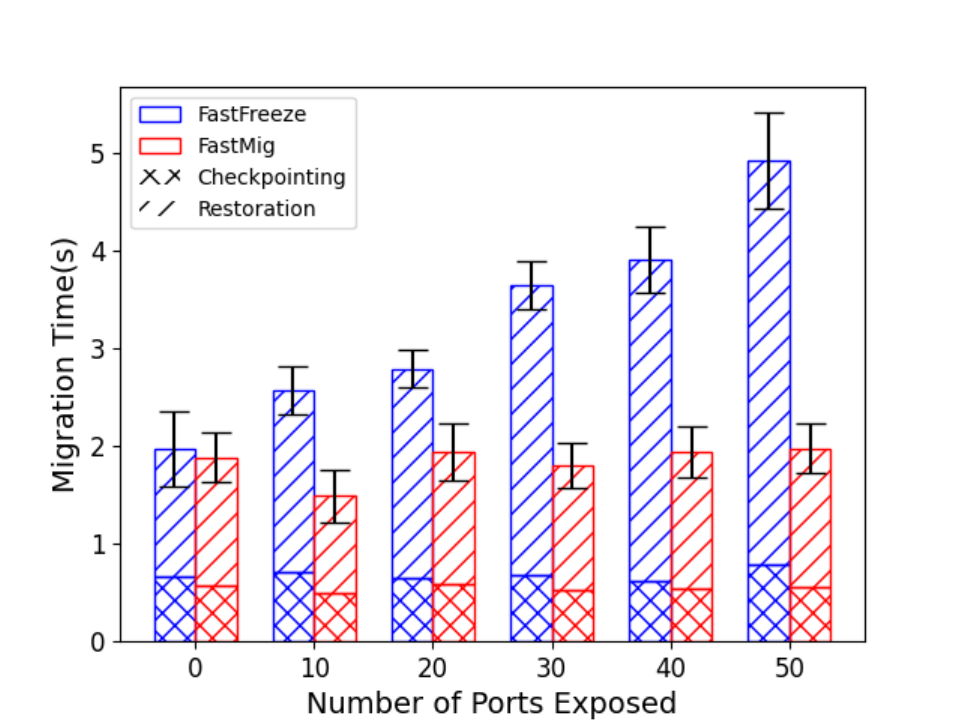}}
    \label{fig:warm_yolo}
    \caption{Migration time of \memhog and \yolo when disabling and enabling warm-restoration. Error bars show 95\%  confidence intervals.}
    \label{fig:migration_time}
\end{figure}

\subsection{Measuring the Overhead of Live Migration}
This experiment aims to measure the migration time overhead of live migration using \ff and, second, with \fastmig that allows migration with a warm restoration at the destination. Both applications were configured for this purpose, \memhog and \yolo.  
To emphasize the impact of the warm restoration, which improves the container migration time by omitting the container startup time at the destination, we measured the migration time on the different numbers of ports exposed on the container since the number of ports exposed is a factor that affects the container startup time \cite{straesser2023empirical}. Each test was conducted 30 times, measuring the total time from the checkpoint phase until the completion of restoration at the destination. The average time was reported in Figure~\ref{fig:migration_time}. 

From the results, when not using \fastmig and not allowing a warm restoration at the destination, the migration time increased linearly as the number of ports increased. Unlike using \fastmig and warm restoration, the migration time was not increased significantly, \ie $\approx$2 seconds. The checkpointing duration among these cases is less affected than the restoration time. 
This shows the benefit of warm restoration. 
The same cases also happened when using \yolo as the evaluated application. 
Although exposing 50 ports is an uncommon use case, it emphasizes the impact of hiding the container startup time. Furthermore, even with only 10 ports exposed, \fastmig demonstrated an approximate 1-second improvement which can already be significant for live migration use cases. It is important to note that not only does exposing ports affect container startup time, but other factors such as the image size and the number of image layers also contribute to the overall startup duration \cite{straesser2023empirical}.

Additionally, we observed that the migration time when equipped with \fastmig is inconsistent. We also observed that the checkpointing time is even between each case, but the restoration time is not. As the restoration operation relies on networking between nodes (\eg operation request across nodes, reading checkpoint from network file system), we surmise that the migration time inconsistency happens due to the network consistency.

\vspace{3mm}
\noindent
\colorbox{blue!10}{
\parbox{0.47\textwidth}{
\textbf{\underline{Takeaway}:} \emph{\fastmig, which enables warm restoration, significantly reduces migration time during live migration by mitigating container startup time.
}
}
}

\begin{figure}[htpb]
    \centerline{\includegraphics[width=0.49\textwidth]{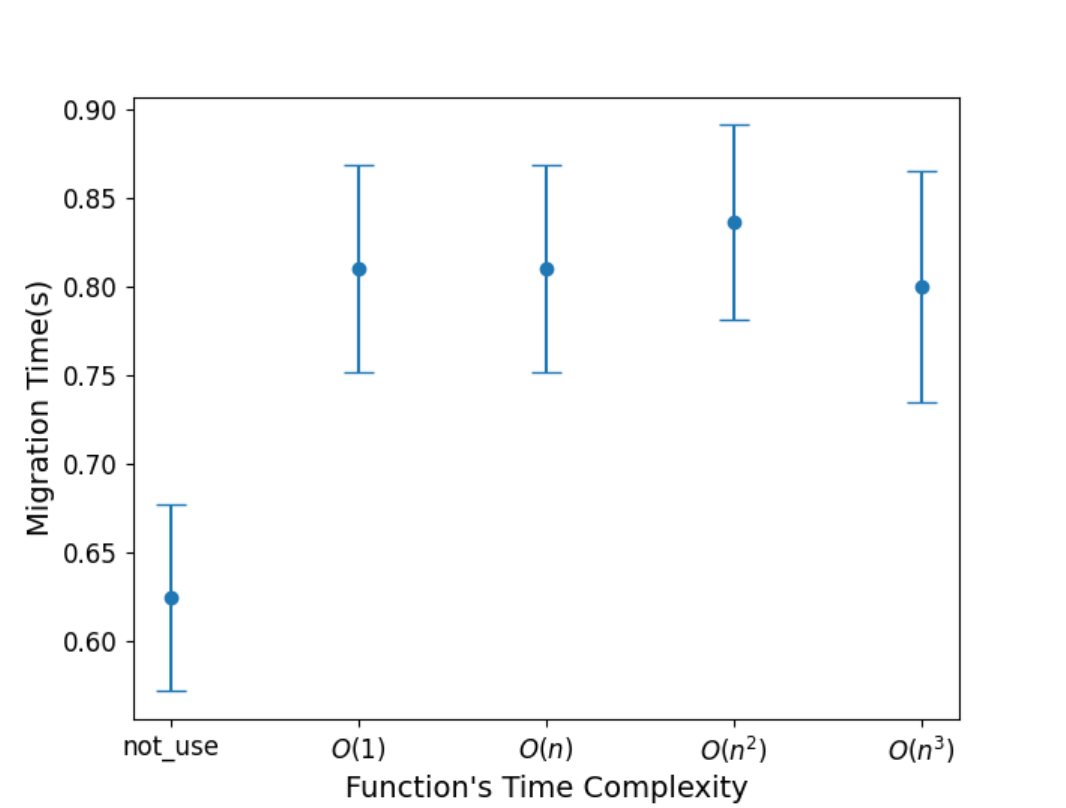}}
    \caption{The impact of embedding the fault tolerance mechanism into the container on the service migration time. Error bars show 95\% confidence intervals.}
    \label{fig:fault_impact}
\end{figure}
\subsection{Impact of the Fault Tolerance Mechanism}
Our proposed solution relies on the logging mechanism and we built a configurable function (used in the Fault Handling Module) to decide on how the container should restart. This experiment focuses on evaluating the overhead of this mechanism by using different string processing logic (function) with different time complexity consisting of $O(1)$, $O(n)$, $O(n^2)$, $O(n^3)$ and also a case that did not use the fault tolerance mechanism. We define the input size (n) as the number of lines from the service log that we feed in. We utilize a 2000-line Apache Web Server log as a simulated input log \cite{zhu2023loghub}.
The experiments were done with 30 repetitions for each function time complexity.

As shown in Figure~\ref{fig:fault_impact}, the fault tolerance mechanism introduces a negligible additional migration overhead of $\approx$0.2 seconds. The overheads are nearly constant across function time complexity as the size of n is small; however, thousand-of-line logs generally suffice to determine the fault cause and the appropriate Start Option Config.


\vspace{3mm}
\noindent
\colorbox{blue!10}{
\parbox{0.47\textwidth}{
\textbf{\underline{Takeaway}:} \emph{\fastmig fault tolerance mechanism with basic log processing introduces little to no impact on migration time.
}
}
}

\subsection{Coverage of the Fault Tolerance Mechanism}
To determine how the fault tolerance mechanism reacts to various types of faults, we set up a fault injection experiment that intentionally causes common faults during each phase of container migration and inspected its actions. To provide generality over various specific use cases, we use the default logic, shown in Listing \ref{lst:default_fault}, that determines \ff restart state from \ff exit code in this experiment. Each fault was injected ten times in each situation where possible.
 
\begin{table*}[htbp]
    \centering
    \begin{NiceTabular}{c || p{2.7 cm} | p{3.5 cm} | p{3.5 cm} }[cell-space-limits=1mm] 
 \textbf{Faults} & \Block{}{\textbf{Normal operation}} & \Block{}{\textbf{Application checkpointing}} & \Block{}{\textbf{Application restoration}} \\
 \hline\hline
 \Block{}{Memory exceed} & \RaggedRight \Block{}{Restart in \\ standby mode}	& \Block{}{Checkpoint stop, service continue} & \Block{}{Restoration succeed} \\ 
 \hline
 \Block{}{Storage exceed} & \RaggedRight \Block{}{Restart in from-scratch mode} & \Block{}{Checkpoint stop, service continue} & \Block{}{Not included} \\
 \hline
\Block{}{Signals (\eg SIGINT, SIGKILL)} & \RaggedRight \Block{}{Restart in \\ standby mode} & \RaggedRight \Block{}{Checkpoint stop, restart in standby mode} & \RaggedRight \Block{}{Restart in \\ from-scratch mode} \\
 \hline
\Block{}{Application unexpected exit \\ (\ie exit code $>$0)} & \RaggedRight \Block{}{Restart in from-scratch mode} & \Block{}{Not included} & \Block{}{Not included} \\
 \hline
\Block{}{Corrupted checkpoint files} & \Block{}{Not included} & \Block{}{Not included} & \RaggedRight \Block{}{Restart in \\ from-scratch mode} \\
 \hline
\Block{}{Network failure (cannot reach NFS)} & \Block{}{Not included} & \RaggedRight \Block{}{Checkpoint stop, service continue} & \Block{}{Restoration freeze until network reconnects} \\
 \hline
\Block{}{Underline system/hardware failure} & \RaggedRight \Block{}{Restart from previous config or standby mode} & \RaggedRight \Block{}{Restart from previous config or standby mode} & \RaggedRight \Block{}{Restart from previous config or standby mode} \\
 
\end{NiceTabular}
    \caption{Types of faults that are injected into each situation. Some faults are not included in the experiment since they are impossible to occur during that phase.}
    \label{tab:2}
\end{table*}

If the application and \ff are restarted and still function properly (\eg can be checkpointed/restored or migrated after the incident is fixed), this will be indicated as a valid result. The setup faults are reported in Table~\ref{tab:2}. Some faults are not included in the experiment since they are impossible to occur during that phase.

Most of the results are valid, as shown in \ref{tab:2}, but only the network failure during application restoration is an exception. This happened because \ff waits to read the checkpoint files from the NFS service, which cannot reach its peer due to network disconnection. NFS blocks \ff process indefinitely and waits for reconnection without a timeout. In this situation, users must either manually stop the container or wait until the network reconnects, which causes the NFS service to unblock the process and resume the restoration.  

Additionally, we observed two limitations of the fault tolerance mechanism, though the results are valid. \revised{Firstly, if a fault occurs during service checkpointing or restoration, the progress of the interrupted operation is discarded. Then, the next attempt of the same operation, \eg a second restoration from the same checkpoint files, is started from the beginning as if it has never been run before.} Secondly, For system/hardware failure, the behavior may depend on how the system (\ie operating system) acts on the container. For instance, when the system is shut down gracefully and sends termination signals to the processes in the container, the Fault Handling Module will control \ff restart mode to be a standby mode, but, in another case, when there is no signal sent to the container (\ie hard shutdown), the logic does not have any chance to evaluate the fault. As a result, the restart mode will be standby as the default mode or will be in the state indicated in the Startup Option Config from the previous function trigger.
 
\vspace{3mm}
\noindent
\colorbox{blue!10}{
\parbox{0.47\textwidth}{
\textbf{\underline{Takeaway}:} \emph{The fault tolerance mechanism enhances the robustness by making the failed container restart at the desired state on the host where the container resides and the failure happens.
}
}
}

\section{Related Works}
\label{sec:related-works}

\revised{
Several studies point out the essential of service liquidity in modern computing architecture; that is, services should be able to run and move freely across underlying platforms. Iorio \etal \cite{liqo} introduced the concept of Liquid Computing, a computing paradigm that abstracts services from the underlying computing continuum. Gallidabino \etal \cite{gallidabino2017architecting} proposed the Liquid Software paradigm that offers a seamless experience to users while migrating across devices. Galantino \etal \cite{galantino2023assessing} addresses the advantages of distributing a fluid workload across diverse computing devices, with a specific focus on the power consumption within the computing continuum.
}

\revised{
Live container migration has been increasingly studied as an alternative solution to VM migration.
Nadgowda \etal \cite{nadgowda2017voyager} presented Voyager, a CRIU-based container migration service that utilizes post-copy filesystem replication. Voyager lazy replication transfers the filesystem of a container in the background when needed while allowing the container to resume operation instantly on the target host, providing zero-downtime data migration and reducing network overhead. 
Benjaponpitak \etal \cite{benjaponpitak2020enabling} proposed CloudHopper, an automated live migration solution across multi-cloud while maintaining connectivity to the client service. CloudHopper allows live migration efficiently through pre-copy techniques and redirects the traffic between the cloud using HAProxy with the cloud provider's VPN. Among others, CloudHopper is the only live migration solution that takes connectivity between computing systems into account.
Ma \etal \cite{ma2018efficient} proposed a framework for offloading services across the edge servers by leveraging the pre-copy live migration technique and eliminating the transfer of redundant container storage layers.
The aforementioned works mainly relied on checkpoint/restore features supported by the container runtime. On the contrary, Souza \etal \cite{junior2022good} presented MyceDrive, a solution to migrate containers within a Kubernetes cluster by embedding checkpoint/restore libraries into the container. It allows only migration of the containerized service memory. \fastmig provides an analogous live migration solution while simultaneously offering robustness enhancement to the container. \fastmig is also open to integration with other migration solutions components through standardized interfaces.
}

\section{Conclusion}\label{sec:conclusion}

In this research, we leverage \ff to establish a robust service liquidity solution, called \fastmig, for the next generation of Cloud computing systems. The main idea behind \fastmig is to separate the service management from its execution. It allows restarting the service without recreating the container and allows post-checkpointing operations, such as those for graceful shutdown instructions, thereby making it robust against failures that occur during the container migration. The decoupled pre-restoration operations spur a warm restoration technique that significantly reduces the overall live migration time. In addition, \fastmig provides an HTTP-based interface; thus, the migration can be requested from the outer component without requiring the privilege to access the container command-line interface, and the solution can be easily integrated with existing systems. 
\fastmig includes a self-feedback fault tolerance mechanism that executes a configurable function to decide how the service should be restarted upon failure occurrence to enhance the robustness. Last but not least, \fastmig is able to efficiently perform migration for multi-process services, commonly needed for service migration, via enhancing the restoration process of such services.
The evaluations show that, firstly, incorporating \ff or \fastmig into the container imposes little to no additional resource and performance overheads in normal service operations(no migration). Secondly, \ff-based live migration solution can operate efficiently with multi-process services when certain privileges are granted to the container through the appropriate container configuration.
Thirdly, the container startup time can be overlapped during the migration when applying the warm restoration technique introduced by \fastmig, which significantly reduces migration time during live migration.
Lastly, we investigated the impact of the self-feedback fault tolerance mechanism and noticed that it introduces a negligible overhead while allowing the handling of flexible types of faults via a simple configuration. 

There are several avenues to extend this research in the future. 
Firstly, machine learning techniques can be added  to enable the flexible fault tolerance mechanism for the undefined faults. Currently, \fastmig reacts to the undefined faults with the default behavior---restart in standby mode.
As such, the second avenue for future research can be to enhance the robustness of service liquidity at the inter-system coordination level. For instance, if the hardware that runs the service fails permanently, the service unavailability should be detected, and a new instance of the service should be started/restored at the last location it has traveled to.
The third avenue for future research can be on the security and privacy aspects of container migration, dealing with challenges such as secure container migration through a third-party network provider or to an untrusted destination system, and also dealing with the authorization and encryption challenges of the migrating container.

%
\bibliographystyle{plain} 
\balance
\bibliography{references}

\begin{thebibliography}{10}

\bibitem{CRIU}
{CRIU Main Page}.
\newblock \url{https://criu.org/Main_Page}.
\newblock Accessed on 2023-12-4.

\bibitem{Docker}
{Docker}.
\newblock \url{https://www.docker.com/}.
\newblock Accessed on 2024-3-21.

\bibitem{k8s}
{Kubernetes}.
\newblock \url{https://kubernetes.io/}.
\newblock Accessed on 2024-3-21.

\bibitem{warmstart}
{Operating Lambda: Performance optimization}.
\newblock \url{https://aws.amazon.com/blogs/compute/operating-lambda-performance-optimization-part-1/ }.
\newblock Accessed on 2024-3-30.

\bibitem{benjaponpitak2020enabling}
Thad Benjaponpitak, Meatasit Karakate, and Kunwadee Sripanidkulchai.
\newblock Enabling live migration of containerized applications across clouds.
\newblock In {\em Proceedings of the 2020-IEEE Conference on Computer Communications}, pages 2529--2538. IEEE, 2020.

\bibitem{chanikaphon2023ums}
Thanawat Chanikaphon and Mohsen~Amini Salehi.
\newblock Ums: Live migration of containerized services across autonomous computing systems.
\newblock In {\em Proceedings of the IEEE Global Communications Conference}, pages 467--472. IEEE, 2023.

\bibitem{denninnart2024smse}
Chavit Denninnart and Mohsen Amini~Salehi.
\newblock Smse: A serverless platform for multimedia cloud systems.
\newblock {\em Concurrency and Computation: Practice and Experience}, 36(4):e7922, 2024.

\bibitem{denninnart2023efficiency}
Chavit Denninnart, Thanawat Chanikaphon, and Mohsen Amini~Salehi.
\newblock Efficiency in the serverless cloud paradigm: A survey on the reusing and approximation aspects.
\newblock {\em Software: Practice and Experience}, 53(10):1853--1886, 2023.

\bibitem{docker-container}
{Docker}.
\newblock What is a container?
\newblock \url{https://www.docker.com/resources/what-container/}.
\newblock Accessed on 2024-3-28.

\bibitem{ermolenko2021internet}
Daniil Ermolenko, Claudia Kilicheva, Ammar Muthanna, and Abdukodir Khakimov.
\newblock {Internet of Things Services Orchestration Framework Based on Kubernetes and Edge Computing}.
\newblock In {\em Proceedings of the IEEE Conference of Russian Young Researchers in Electrical and Electronic Engineering (ElConRus)}, pages 12--17, 2021.

\bibitem{galantino2023assessing}
Stefano Galantino, Fulvio Risso, Vlad~C Coroam{\u{a}}, and Antonio Manzalini.
\newblock {Assessing the Potential Energy Savings of a Fluidified Infrastructure}.
\newblock {\em Computer}, 56(6):26--34, 2023.

\bibitem{gallidabino2017architecting}
Andrea Gallidabino, Cesare Pautasso, Tommi Mikkonen, Kari Syst{\"a}, Jari-Pekka Voutilainen, and Antero Taivalsaari.
\newblock {Architecting Liquid Software.}
\newblock {\em J. Web Eng.}, 16(5\&6):433--470, 2017.

\bibitem{ghatrehsamani2020art}
Davood Ghatrehsamani, Chavit Denninnart, Josef Bacik, and Mohsen Amini~Salehi.
\newblock The art of cpu-pinning: Evaluating and improving the performance of virtualization and containerization platforms.
\newblock In {\em Proceedings of the 49th International conference on parallel processing}, pages 1--11, 2020.

\bibitem{imran2020multi}
Hamza~Ali Imran, Usama Latif, Ataul~Aziz Ikram, Maryam Ehsan, Ahmed~Jamal Ikram, Waleed~Ahmad Khan, and Saad Wazir.
\newblock Multi-cloud: a comprehensive review.
\newblock In {\em Proceedings of the 23rd International Multitopic Conference (INMIC 2020)}, pages 1--5. IEEE, 2020.

\bibitem{liqo}
Marco Iorio, Fulvio Risso, Alex Palesandro, Leonardo Camiciotti, and Antonio Manzalini.
\newblock {Computing Without Borders: The Way Towards Liquid Computing}.
\newblock {\em IEEE Transactions on Cloud Computing}, 11(3):2820--2838, 2023.

\bibitem{kim2021resource}
Eunsook Kim, Kyungwoon Lee, and Chuck Yoo.
\newblock {On the Resource Management of Kubernetes}.
\newblock In {\em Proceedings of the International Conference on Information Networking (ICOIN)}, pages 154--158, 2021.

\bibitem{ma2018efficient}
Lele Ma, Shanhe Yi, Nancy Carter, and Qun Li.
\newblock Efficient live migration of edge services leveraging container layered storage.
\newblock {\em IEEE Transactions on Mobile Computing}, 18(9):2020--2033, 2018.

\bibitem{mancuso2023efficiency}
Vincenzo Mancuso, Leonardo Badia, Paolo Castagno, Matteo Sereno, and Marco~Ajmone Marsan.
\newblock Efficiency of distributed selection of edge or cloud servers under latency constraints.
\newblock In {\em Proceedings of the 21st Mediterranean Communication and Computer Networking Conference (MedComNet 2023)}, pages 158--166. IEEE, 2023.

\bibitem{nadgowda2017voyager}
Shripad Nadgowda, Sahil Suneja, Nilton Bila, and Canturk Isci.
\newblock Voyager: Complete container state migration.
\newblock In {\em Proceedings of the 37th International Conference on Distributed Computing Systems (ICDCS)}, pages 2137--2142. IEEE, 2017.

\bibitem{CAPCHECKREST}
Adrian Reber.
\newblock capabilities: Introduce {CAP\_CHECKPOINT\_RESTORE}.
\newblock \url{https://patchwork.kernel.org/project/linux-security-module/patch/20200715144954.1387760-2-areber@redhat.com/}, 2020.
\newblock Accessed on 2023-12-4.

\bibitem{redhat-containers}
{Red Hat}.
\newblock Understanding containers.
\newblock \url{https://www.redhat.com/en/topics/containers}.
\newblock Accessed on 2024-3-28.

\bibitem{redmon2018yolov3}
Joseph Redmon and Ali Farhadi.
\newblock Yolov3: An incremental improvement.
\newblock {\em arXiv preprint arXiv:1804.02767}, 2018.

\bibitem{singh2021taxonomy}
Gursharan Singh and Parminder Singh.
\newblock A taxonomy and survey on container migration techniques in cloud computing.
\newblock {\em Sustainable Development Through Engineering Innovations: Select Proceedings of SDEI 2020}, pages 419--429, 2021.

\bibitem{junior2022good}
Paulo Souza~Junior, Daniele Miorandi, and Guillaume Pierre.
\newblock Good shepherds care for their cattle: Seamless pod migration in geo-distributed kubernetes.
\newblock In {\em Proceedings of the 6th IEEE International Conference on Fog and Edge Computing (ICFEC)}, pages 26--33. IEEE, 2022.

\bibitem{stoyanov2018efficient}
Radostin Stoyanov and Martin~J Kollingbaum.
\newblock Efficient live migration of linux containers.
\newblock In {\em Proceedings of the ISC High Performance 2018 International Workshops, Frankfurt/Main, Germany, June 28, 2018, Revised Selected Papers 33}, pages 184--193. Springer, 2018.

\bibitem{straesser2023empirical}
Martin Straesser, Andr{\'e} Bauer, Robert Leppich, Nikolas Herbst, Kyle Chard, Ian Foster, and Samuel Kounev.
\newblock {An empirical study of container image configurations and their impact on start times}.
\newblock In {\em Proceedings of the 23rd International Symposium on Cluster, Cloud and Internet Computing (CCGrid)}, pages 94--105. IEEE, 2023.

\bibitem{ullah2022fedfly}
{Ullah, Rehmat, Di Wu, Paul Harvey, Peter Kilpatrick, Ivor Spence, and Blesson Varghese}.
\newblock {FedFly: Toward migration in edge-based distributed federated learning}.
\newblock {\em IEEE Communications Magazine}, 60(11):42--48, 2022.

\bibitem{FastFreeze}
Nicolas Viennot.
\newblock {FastFreeze}.
\newblock \url{https://github.com/twosigma/fastfreeze}.
\newblock Accessed on 2023-11-20.

\bibitem{zhu2023loghub}
Jieming Zhu, Shilin He, Pinjia He, Jinyang Liu, and Michael~R Lyu.
\newblock Loghub: A large collection of system log datasets for ai-driven log analytics.
\newblock In {\em Proceedings of the 34th International Symposium on Software Reliability Engineering (ISSRE)}, pages 355--366. IEEE, 2023.

\end{thebibliography}


\end{document}